\documentclass[lettersize,journal]{IEEEtran}

\usepackage{amsmath,amsfonts,amssymb}
\usepackage{algorithmic}
\usepackage{algorithm}
\usepackage{array}
\usepackage[caption=false,font=normalsize,labelfont=sf,textfont=sf]{subfig}
\usepackage{textcomp}
\usepackage{stfloats}
\usepackage{url}
\usepackage{verbatim}
\usepackage{graphicx}
\usepackage{cite}
\usepackage{booktabs}
\usepackage{microtype}
\usepackage{xspace}
\usepackage{bm}

\usepackage{xcolor}
\usepackage[
  colorlinks=true,
  linkcolor=blue,   
  citecolor=blue,   
  urlcolor=black
]{hyperref}

\usepackage[capitalize]{cleveref}

\Crefname{section}{Section}{Sections}
\Crefname{figure}{Fig.}{Figs.}
\Crefname{table}{Table}{Tables}

\newcommand{\method}{ODMA\xspace}
\newcommand{\racm}{RACM\xspace}

\begin{document}

\title{\method: On-Demand Memory Allocation Strategy for LLM Serving on LPDDR-Class Accelerators}

\author{Guoqiang Zou, Wanyu Wang, Hao Zheng, Longxiang Yin, and Yinhe Han
\thanks{Manuscript received [Date]; revised [Date]. (Corresponding author: Longxiang Yin.)}
\thanks{Guoqiang Zou and Wanyu Wang are with the University of Chinese Academy of Sciences, Beijing, China. (e-mail: zouguoqiang23@mails.ucas.ac.cn; wangwanyu23@mails.ucas.ac.cn).}
\thanks{Hao Zheng is with Beijing Information Science and Technology University. (e-mail: zhenghao127@bistu.edu.cn).}
\thanks{Longxiang Yin and Yinhe Han are with the Institute of Computing Technology, Chinese Academy of Sciences. (e-mail: yinlongxiang@ict.ac.cn; yinhes@ict.ac.cn).}
}


\maketitle

\begin{abstract}
Existing memory management techniques severely hinder efficient Large Language Model (LLM) serving on accelerators constrained by poor random-access bandwidth. While static pre-allocation preserves memory contiguity, it incurs significant overhead due to worst-case provisioning. Conversely, fine-grained paging mitigates this overhead but relies on HBM's high random-access tolerance, making it unsuitable for LPDDR systems where non-sequential access rapidly degrades bandwidth. Furthermore, prior works typically assume static distributions and HBM characteristics, thereby failing to resolve the critical fragmentation and bandwidth constraints inherent to LPDDR hardware.

We present \method, an \emph{on-demand memory allocation} strategy tailored for \emph{random-access-constrained} (\racm) accelerators, such as the Cambricon MLU series. \method advances generation-length prediction by addressing two critical limitations in production workloads: (i) distribution drift that invalidates static bucket boundaries, and (ii) performance fragility under heavy-tailed request patterns. \method integrates a lightweight length predictor with \emph{adaptive bucket partitioning} and a \emph{fallback safety pool}. Bucket boundaries are dynamically recalibrated via online histograms to maximize utilization, while the safety pool ensures robustness against prediction errors. On Alpaca and Google-NQ benchmarks, \method improves S3's prediction accuracy from 98.60\% to 99.55\% and 82.68\% to 93.36\%, respectively. Deployment with DeepSeek-R1-Distill-Qwen-7B on Cambricon MLU370-X4 accelerators demonstrates that \method increases KV-cache utilization by up to 19.25\% (absolute) and throughput (TPS) by 23--27\% over static baselines, validating the efficacy of predictor-driven contiguous allocation for LPDDR-class devices.
\end{abstract}


\section{Introduction}
\IEEEPARstart{L}{arge}-scale Transformer models (LLMs)~\cite{brown2020gpt3,vaswani2017attention,devlin2018bert} have become central to modern intelligent applications. In production environments, inference throughput is typically bottlenecked by memory capacity and bandwidth rather than raw compute capability. This limitation arises from autoregressive decoding, which maintains a Key-Value (KV) cache whose footprint scales linearly with sequence length and batch size, consuming a substantial fraction of device memory~\cite{pope2023efficient,s3,pagedattention}. As a result, KV-cache management directly governs admission control: conservative provisioning limits concurrency, whereas under-provisioning risks costly recomputation and system thrashing. Modern serving stacks therefore prioritize memory efficiency, commonly relying on static reservation or paging-based techniques~\cite{s3,zhou2024survey}.

On HBM-equipped GPUs, paging-based architectures (e.g., PagedAttention~\cite{pagedattention}) effectively reduce fragmentation by decoupling logical sequence layout from physical memory placement. These designs critically depend on HBM’s ability to sustain high bandwidth under irregular, non-contiguous access patterns. In sharp contrast, a large class of production accelerators—exemplified by the Cambricon MLU370 series~\cite{cambriconmlu370x4}—employs LPDDR5-class memory. On such Random Access Costly Memory (\racm{}) architectures, fine-grained random access incurs a severe bandwidth penalty relative to sequential streaming~\cite{steiner2022lpddr5}. Paging further amplifies this bottleneck by scattering KV fetches across memory, significantly degrading effective bandwidth. Conversely, traditional static pre-allocation preserves contiguity and streaming-friendly access, but suffers from low utilization due to the high variance in request lengths.

\subsection{Problem Statement}
LLM workloads exhibit generation lengths with significant variance and heavy-tailed distributions. Interactive queries often terminate quickly, while reasoning-intensive tasks produce long outputs. Moreover, production traces show that length distributions drift over time as user behavior and prompt templates evolve~\cite{s3,pagedattention}. Under these conditions, worst-case provisioning drastically reduces admitted concurrency, whereas aggressive over-commitment risks memory overflow. A practical allocator for \racm{} accelerators must therefore satisfy two conflicting requirements: it must be \emph{length-aware} to minimize memory waste, yet \emph{layout-conscious} to enforce contiguous access patterns and preserve bandwidth efficiency.

\subsection{Limitations of Prior Art}
S3~\cite{s3} introduced length prediction with bucketed scheduling, but relies on static bucket boundaries tuned for HBM-based systems. In dynamic production environments, these static boundaries fail to adapt to distribution drift, leading to substantial internal fragmentation even when predictions are accurate. PagedAttention~\cite{pagedattention} eliminates external fragmentation via block-level management, but presumes hardware tolerance for random access—an assumption that does not hold on LPDDR-based systems. Similarly, kernel-level optimizations such as FlashInfer~\cite{FlashInfer} target HBM-equipped GPUs and do not address the allocator-level bandwidth penalties inherent to paging on \racm{} devices.

\subsection{Contribution: \method}
We propose \method, a predictor-driven, hardware-aware allocation strategy for \racm{} accelerators. \method reconciles memory efficiency with bandwidth constraints through two complementary mechanisms: (i) \textbf{adaptive bucketing} derived from live production traces to mitigate distribution drift, and (ii) a \textbf{fallback safety mechanism} (the Large Bucket) that trades marginal memory capacity for robustness against long-tail mispredictions. Prior to decoding, \method predicts generation length, allocates a matched contiguous memory block, and routes high-uncertainty requests to the safety pool. This design preserves the streaming-friendly access patterns required by LPDDR while significantly improving memory utilization, offering a practical alternative to paging for bandwidth-constrained architectures.

\section{Background and Motivation}

\subsection{LLM Inference and Memory Bottlenecks}
Transformer-based inference requires maintaining a dynamically growing per-request KV-cache. As sequence length increases, this cache rapidly dominates overall memory consumption, thereby constraining the maximum achievable batch size (i.e., concurrency) and shaping the memory traffic profile of inference workloads. Paging-based designs (e.g., PagedAttention~\cite{pagedattention}) have become the de facto standard on HBM-equipped GPUs, as they enable flexible, non-contiguous allocation and alleviate fragmentation. However, this flexibility comes at the cost of memory access indirection, whose performance impact is highly dependent on the underlying hardware architecture.

\subsection{Random-Access-Constrained Memory (\racm)}
LPDDR5-class memory, widely deployed in cost-effective accelerators, exhibits a pronounced performance asymmetry between sequential and random access patterns. While sequential (streaming) traffic can approach peak bandwidth, characterization studies show that small-granularity random accesses severely degrade throughput, often limiting effective bandwidth to approximately $66\%$ of peak even under ideal conditions~\cite{steiner2022lpddr5}. This degradation arises from reduced burst efficiency and increased bank conflicts inherent to random traffic. As a result, the irregular access patterns induced by fine-grained paging become a primary throughput bottleneck on LPDDR-based systems.

We define a device as \emph{random-access-constrained} (\racm) if the ratio of sustained random-access bandwidth ($B_{\text{rand}}$) to sequential-access bandwidth ($B_{\text{seq}}$) satisfies:
\begin{equation}
\frac{B_{\text{rand}}}{B_{\text{seq}}} \lesssim \alpha < 1,
\end{equation}
where $\alpha$ typically lies in the range of $\approx 0.3$--$0.7$ for LPDDR4/5 configurations. In this regime, preserving contiguous allocation is critical for achieving high efficiency, yet conventional static allocation strategies incur severe memory inefficiency under variable-length workloads. \method resolves this fundamental tension by dynamically matching contiguous memory reservations to predicted demand.

\section{\method Design}
\label{sec:design}

\method is a predictor-driven strategy tailored for \racm accelerators. It preserves the LPDDR-friendly contiguous memory layout while eliminating the waste of static provisioning. \method achieves this via three mechanisms: predictive reservation, adaptive bucket recalibration, and a fallback safety pool.

\subsection{High-Level Architecture}
\begin{figure}[!t]
  \centering
  \includegraphics[width=\columnwidth]{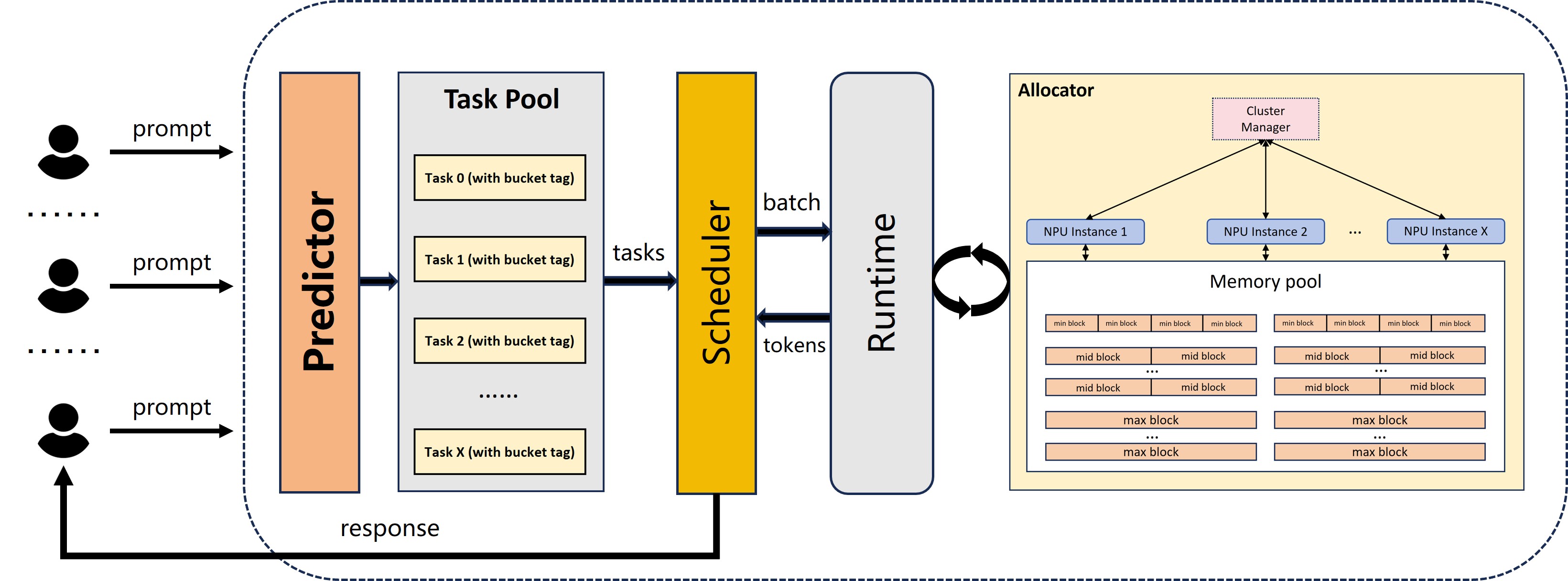}
  \caption{\method overview: Predictor annotates prompts; Scheduler groups tasks; Allocator manages contiguous LPDDR blocks.}
  \label{fig:design}
\end{figure}

\Cref{fig:design} depicts the pipeline. The \textbf{Predictor} assigns a length estimate $\hat{L}$ and bucket tag to requests. The \textbf{Scheduler} batches compatible tasks to minimize padding. The \textbf{Allocator} manages device-local free lists, dispensing \emph{contiguous} blocks to ensure attention kernels utilize peak streaming bandwidth. Upon completion, the \textbf{Runtime} asynchronously logs realized lengths to refresh bucket boundaries.

\subsection{Uncertainty-Aware Prediction}
\begin{figure}[!t]
  \centering
  \includegraphics[width=0.9\columnwidth]{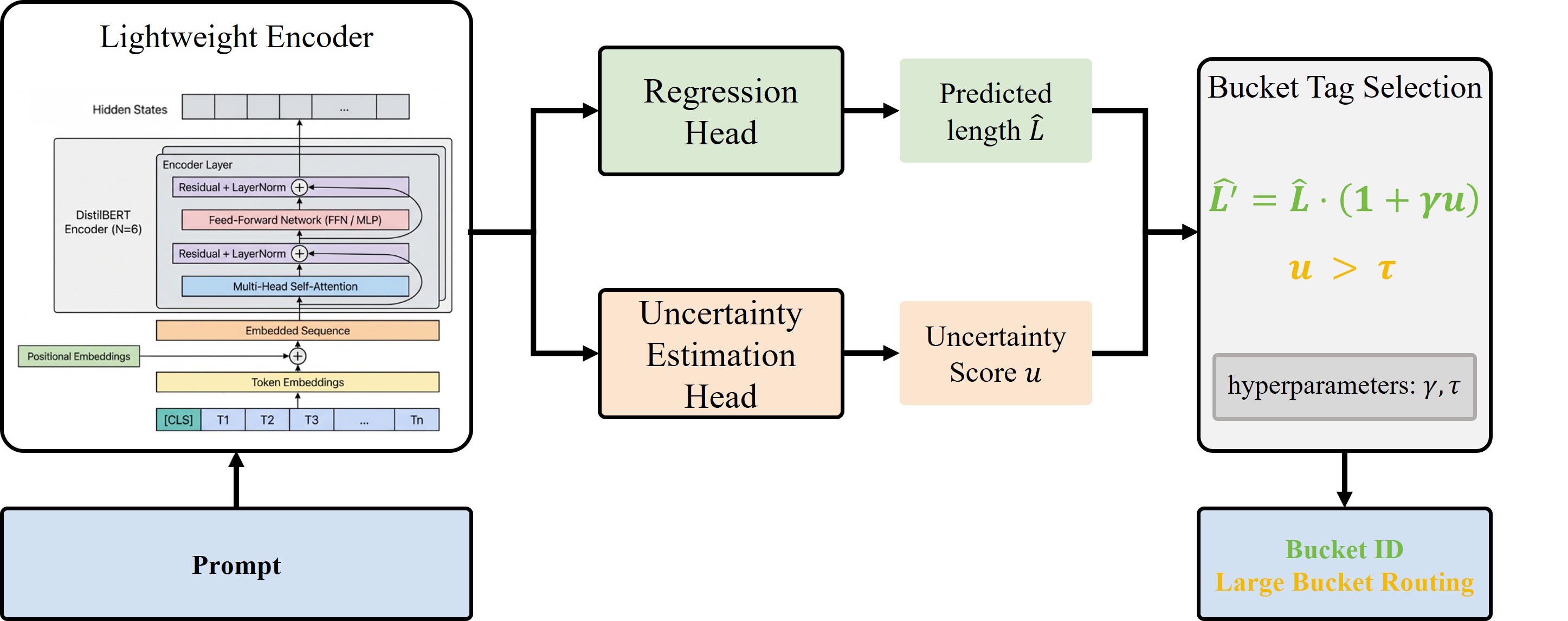}
  \caption{Predictor architecture and uncertainty-aware routing. A lightweight encoder outputs a length estimate $\hat{L}$ and uncertainty $u$; the estimate is inflated to $\hat{L}'=\hat{L}(1+\gamma u)$, and high-uncertainty requests ($u>\tau$) are routed to the Large Bucket.}
  \label{fig:predictor}
\end{figure}

We employ a lightweight encoder to predict length $\hat{L}$ and uncertainty $u$ (\Cref{fig:predictor}). To decouple common-case efficiency from tail safety, we inflate estimates via $\hat{L}' = \hat{L} \cdot (1 + \gamma u)$. Requests with high uncertainty ($u > \tau$) represent risk and are routed to a fallback pool. In our implementation ($\gamma=0.2, \tau=0.8$), this policy maintains tight buckets for the majority of traffic while explicitly budgeting for uncertainty.

\subsection{Adaptive Bucket Manager}
To counteract distribution drift, \method recalibrates bucket boundaries $\{b_i\}$ using empirical quantiles $Q_{p_i}(L)$ from a sliding window of recent requests:
\begin{equation}
\mathcal{B} = \{b_1, \dots, b_B\}, \quad b_i = Q_{p_i}(L).
\end{equation}
This quantile-based approach automatically aligns buckets with the evolving workload. We use $B=4$ buckets and a 10,000-request window, refreshing boundaries asynchronously every 1,000 requests. This configuration balances adaptivity with stability while keeping metadata overhead negligible.

\subsection{Large-Bucket Safety Strategy}
A dedicated \emph{Large Bucket} provides robustness against tail events without fine-grained paging:
\begin{itemize}
    \item \textbf{Pre-emptive Routing:} High-uncertainty requests ($u > \tau$) bypass standard buckets and allocate directly from the safety pool.
    \item \textbf{Reactive Migration:} If a request overflows its bucket, the runtime triggers a \emph{mid-decode migration}. Since source and destination are contiguous, this sequential copy exploits LPDDR's high streaming bandwidth.
\end{itemize}
Migration is rare ($<0.5\%$), ensuring the safety mechanism incurs minimal throughput overhead.

\subsection{Allocator Integration}
The Allocator exposes standard \texttt{reserve}/\texttt{release} APIs backed by per-size free lists. A background thread handles histogram updates and boundary refreshes to avoid blocking the critical path. In multi-device nodes, a Cluster Manager synchronizes logical bucket configurations to ensure consistent scheduling.

\section{Implementation}
\label{sec:implementation}

\subsection{Hardware and Software Stack}
We deploy \method on a node equipped with four Cambricon MLU370-X4 accelerators. Each card provides 24\,GB of LPDDR5 device memory and 307.2\,GB/s peak bandwidth~\cite{cambriconmlu370x4}. All experiments serve DeepSeek-R1-Distill-Qwen-7B, a contemporary 7B-class model. The runtime is built by extending Cambricon-vLLM, with the predictor and dynamic allocator integrated behind a vLLM-compatible API. The integration is intentionally minimally invasive: the primary modifications are (i) invoking the predictor prior to admission control, and (ii) replacing static KV-cache reservation with bucket-based contiguous allocation, while preserving the original scheduler and batching policies unchanged.

\subsection{Predictor Instantiation}
\begin{figure}[!t]
  \centering
  \includegraphics[width=0.75\columnwidth]{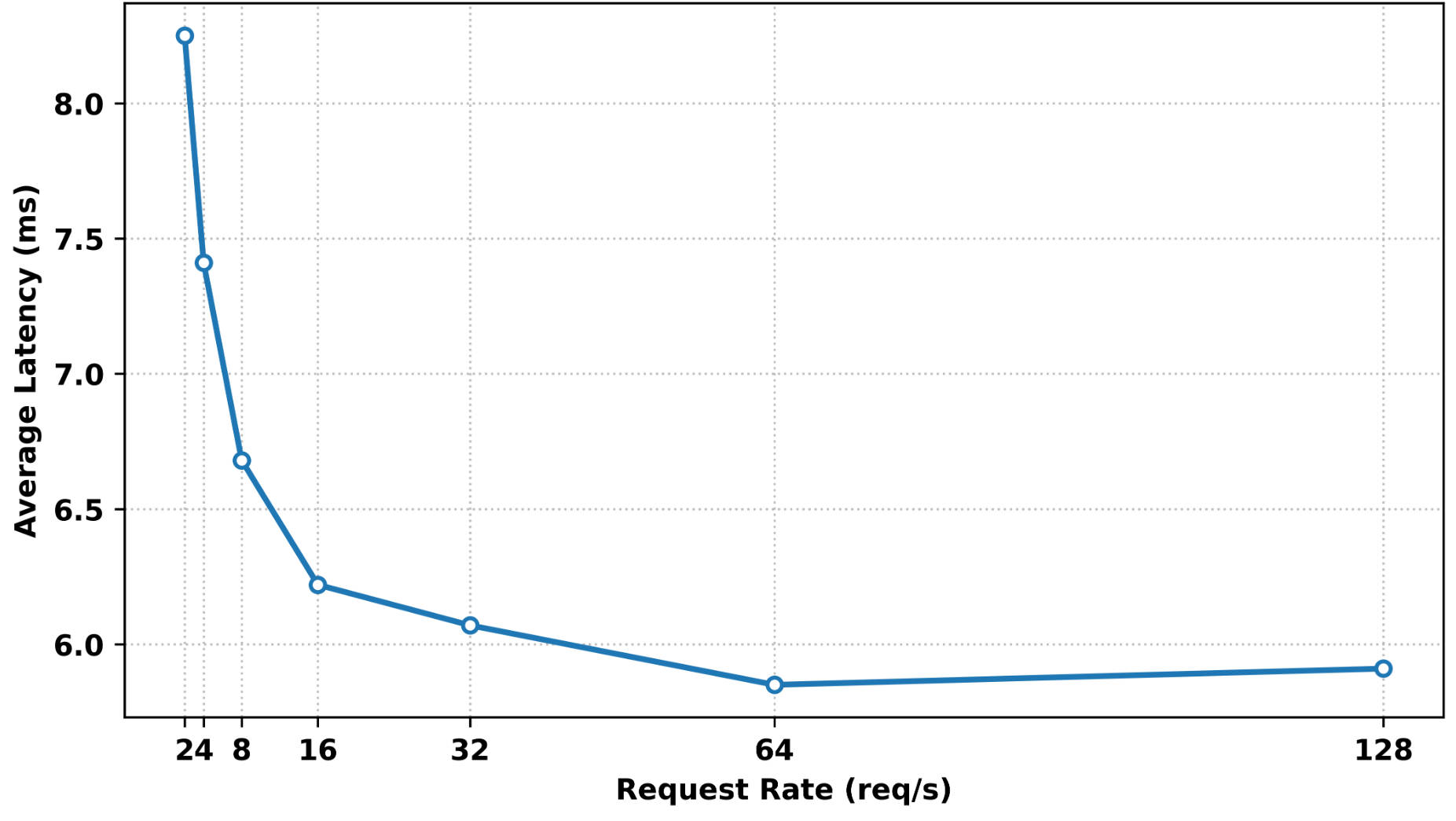}
  \caption{Average predictor inference latency under varying request arrival rates. The predictor uses concurrent execution with micro-batching; higher arrival rates enable more effective batch formation, amortizing per-batch overhead and reducing average per-request predictor latency.}
  \label{fig:predlat}
\end{figure}

We instantiate a DistilBERT-scale encoder with 66M parameters, following the design of S3~\cite{s3}. Predictor inputs include tokenized prompts along with lightweight request metadata. Predictor inference latency is on the order of milliseconds and is negligible relative to end-to-end LLM decoding latency, making the overhead acceptable in practice. We implement a concurrent predictor using micro-batching: incoming requests are opportunistically aggregated into small batches and executed jointly to improve device utilization. As request arrival rates increase, batch formation becomes more consistent, amortizing per-batch overhead and reducing the average \emph{per-request} predictor inference latency, as shown in \Cref{fig:predlat}. 

Unless otherwise specified, the reported predictor latency measures only the predictor execution time—including micro-batch formation and model inference—but excludes LLM prefill and decoding time. This isolates the incremental overhead introduced by \method’s predictor. The predictor outputs both the length estimate $\hat{L}$ and the uncertainty score $u$, enabling risk-aware allocation without frequent retraining; workload drift is instead handled through online updates to bucket boundaries.

\subsection{Allocator Integration}
A background thread continuously consumes instrumentation logs to update length histograms and re-derive bucket boundaries. Boundary updates are applied only to newly admitted requests, ensuring that in-flight decoding is unaffected. Large-bucket migration is implemented via an allocate-and-copy mechanism. In practice, occupancy of the large bucket remains low, effectively trading a small amount of additional memory overhead for robustness under heavy-tailed workloads, while preserving contiguous, streaming-friendly access patterns.

\section{Evaluation}
\label{sec:evaluation}
We evaluate on Alpaca~\cite{taori2023alpaca} and Google-NQ~\cite{kwiatkowski2019natural}. We compare (i) length-prediction accuracy against S3~\cite{s3}, and (ii) throughput/utilization against a static worst-case pre-allocation baseline (Cambricon-vLLM 0.6.2). For each dataset, both systems are driven by the same offered load with Poisson arrivals and process the same number of requests; all serving settings are identical except for the KV-cache allocator (i.e., scheduler and batching are unchanged). We define TPS as average output tokens per second over the entire experiment. During execution, we record for each request the reserved KV-cache size assigned by the allocator and the actual KV-cache usage; after all requests complete, we compute utilization as total actual KV usage divided by total reserved KV capacity across all requests. Unless otherwise stated, predictor latency refers to predictor-side execution time and does not include request queueing delay in the serving system.

\subsection{Prediction Accuracy}
\begin{figure}[!t]
  \centering
  \includegraphics[width=\columnwidth]{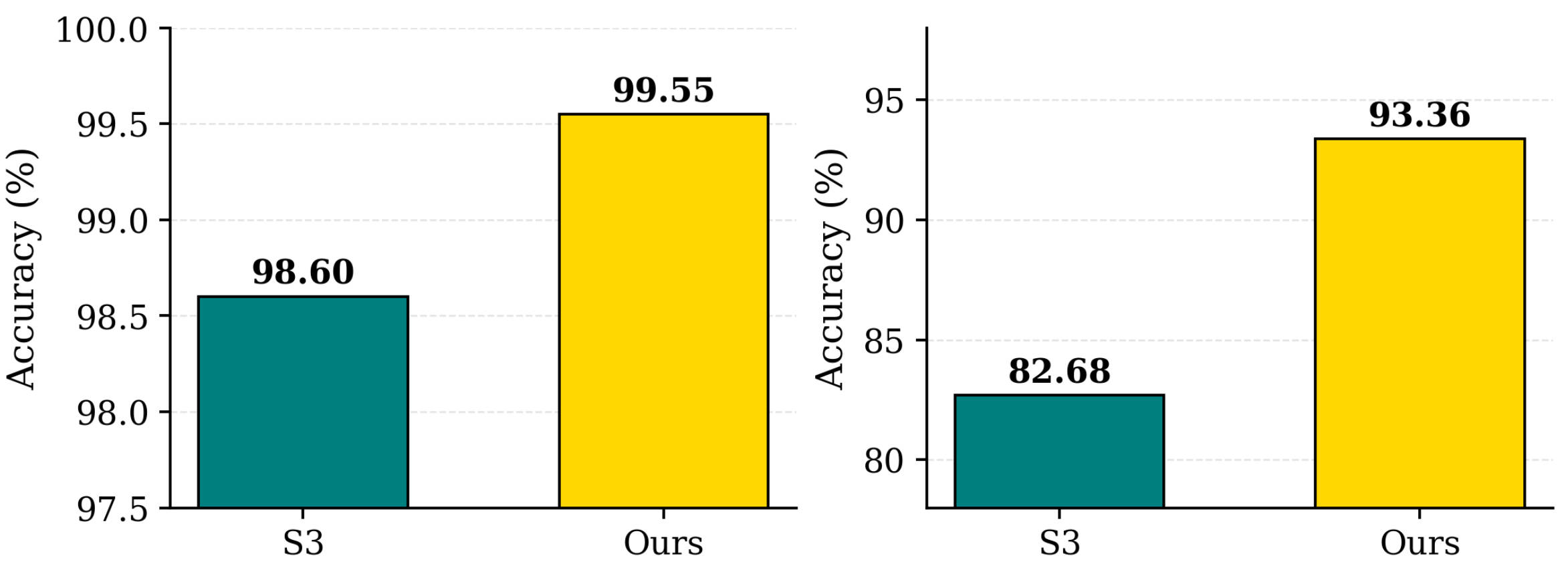}
  \caption{Prediction accuracy: \method vs.\ S3~\cite{s3} on Alpaca (Left) and Google-NQ (Right).}
  \label{fig:accuracy}
\end{figure}

\Cref{fig:accuracy} shows that \method improves prediction accuracy by re-learning boundaries. On Alpaca, accuracy rises from 98.60\% (S3) to 99.55\%; on Google-NQ, from 82.68\% to 93.36\%. These gains are important because tighter buckets only help when under-allocation is rare. Dynamic boundaries reduce systematic mismatch under drift, while uncertainty-aware inflation and the large bucket handle the remaining tail cases.

\subsection{Throughput and Utilization}
\begin{figure}[!t]
  \centering
  \includegraphics[width=\columnwidth]{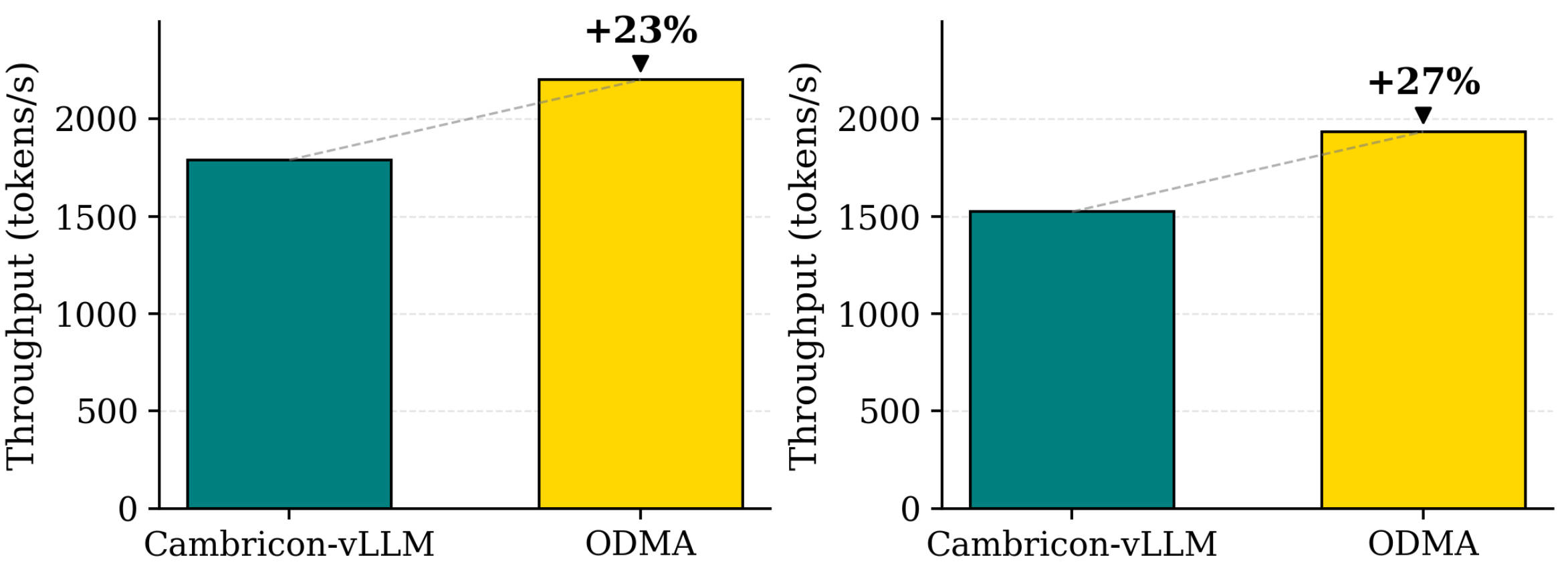}
  \caption{Average output-token throughput (TPS) improvement of \method over the static baseline.}
  \label{fig:throughput}
\end{figure}

\Cref{fig:throughput} reports end-to-end throughput (TPS), measured as average output tokens per second over the experiment. On the MLU370 node, \method improves TPS by 23\% on Alpaca and 27\% on Google-NQ by reducing internal fragmentation and admitting more concurrent requests. Because KV blocks remain contiguous, decoding maintains streaming-friendly access patterns on LPDDR-class devices, avoiding the random-access penalty associated with fine-grained paging.

\begin{figure}[!t]
  \centering
  \includegraphics[width=\columnwidth]{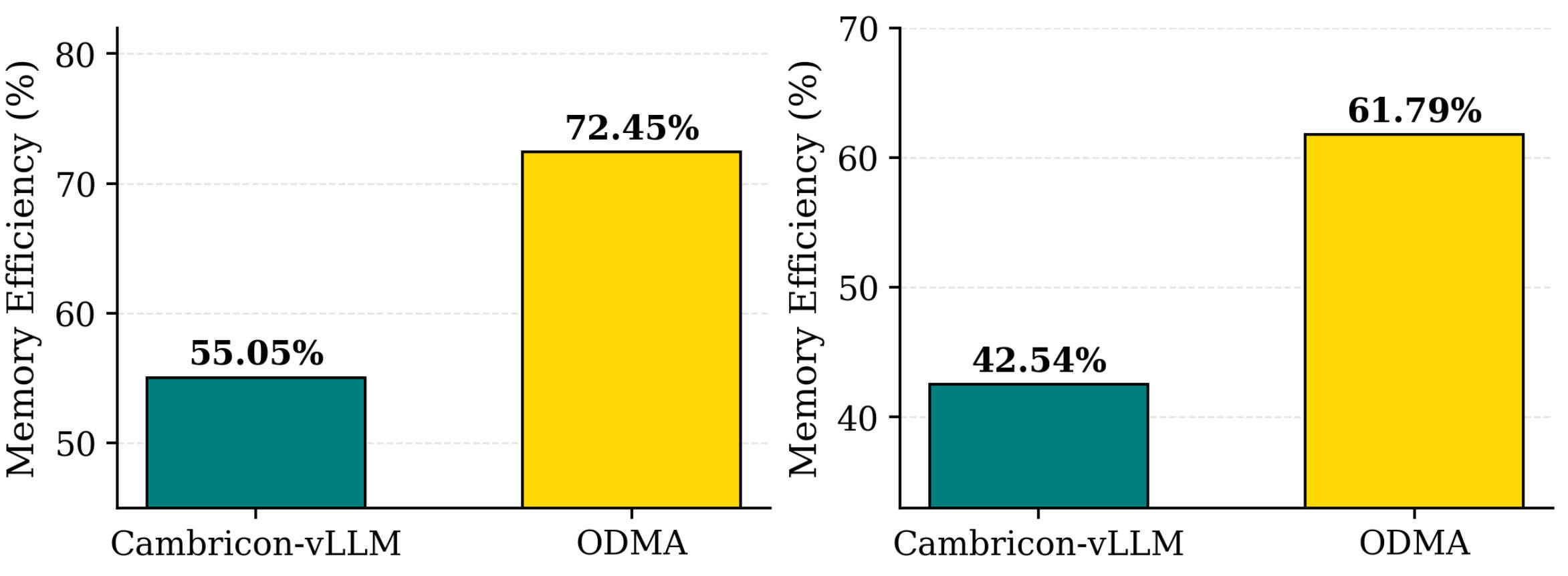}
  \caption{KV-cache utilization ($\sum \text{KV}_{\text{actual}}/\sum \text{KV}_{\text{reserved}}$) with \method. Left: Alpaca; Right: Google-NQ.}
  \label{fig:util}
\end{figure}

\Cref{fig:util} shows KV-cache utilization, computed as $\sum \text{KV}_{\text{actual}} \,/\, \sum \text{KV}_{\text{reserved}}$ across all requests in the experiment. Importantly, the utilization and TPS results are obtained from the same offered load on each dataset: higher utilization indicates less over-reservation of KV capacity, which increases effective headroom and allows the system to sustain higher admitted concurrency, leading to higher output-token throughput.

\section{Conclusion}
\method demonstrates that efficient LLM serving on memory-constrained hardware does not strictly require paging. By leveraging accurate length prediction, adaptive bucketing, and a robust fallback mechanism, \method achieves on-demand contiguous allocation. This approach unlocks substantial improvements in utilization and throughput on LPDDR-based accelerators without requiring kernel-level modifications. Our findings underscore the importance of co-designing allocation policies with memory subsystem characteristics: for hardware where random access is costly, intelligent contiguous allocation offers a superior alternative to indiscriminate paging.

\bibliographystyle{IEEEtran}

\end{document}